\documentstyle [12pt,preprint,aps,amsfonts] {revtex}
\tighten
\draft
\widetext
\preprint{YCTP-P12-99, BU-HEP-99-8, SLAC-PUB-8139, ITP-SB-98-74}
\newcommand{\beq}{\begin{equation}}
\newcommand{\eeq}{\end{equation}}
\newcommand{\beqs}{\begin{eqnarray}}
\newcommand{\eeqs}{\end{eqnarray}}

\newcommand{\drawsquare}[2]{\hbox{%
\rule{#2pt}{#1pt}\hskip-#2pt
\rule{#1pt}{#2pt}\hskip-#1pt
\rule[#1pt]{#1pt}{#2pt}}\rule[#1pt]{#2pt}{#2pt}\hskip-#2pt
\rule{#2pt}{#1pt}}
\newcommand{\fund}{\raisebox{-.5pt}{\drawsquare{6.5}{0.4}}}
\newcommand{\sym}{\raisebox{-.5pt}{\drawsquare{6.5}{0.4}}\hskip-0.4pt%
        \raisebox{-.5pt}{\drawsquare{6.5}{0.4}}}
\newcommand{\asym}{\raisebox{-3.5pt}{\drawsquare{6.5}{0.4}}\hskip-6.9pt%
        \raisebox{3pt}{\drawsquare{6.5}{0.4}}}

\begin{document}

\draft

\baselineskip 6.0mm

\bigskip
\bigskip

\title{ New Constraints on Chiral Gauge Theories}

\vspace{6mm}

\author{Thomas Appelquist$^{(a)}$\thanks{email: thomas.appelquist@yale.edu} 
\and Andrew Cohen$^{(b)}$\thanks{email: cohen@bu.edu} \and
Martin Schmaltz$^{(c)}$\thanks{email: schmaltz@slac.stanford.edu} \and
Robert Shrock$^{(d)}$ \thanks{email: robert.shrock@sunysb.edu}}

\vspace{6mm}

\address{(a) \ Department of Physics, Yale University, New Haven, CT  06511}

\address{(b) \ Department of Physics, Boston University, Boston, MA  02215}

\address{(c) \ SLAC, Stanford, CA  94309}

\address{(d) \ Institute for Theoretical Physics, State University of New
York, Stony Brook, NY 11794}

\maketitle

\begin{abstract}

Recently, a new constraint on the structure of a wide class of strongly coupled
field theories has been proposed. It takes the form of an inequality limiting
the number of degrees of freedom in the infrared description of a theory to be
no larger than the number of underlying, ultraviolet degrees of freedom. Here
we apply this inequality to chiral gauge theories. For some models we find that
it is always satisfied, while for others we find that the assumption of the
validity of the inequality implies a strong additional restriction on the
spectrum of massless composite particles.

\end{abstract}

\vspace{16mm}

\newpage

\pagestyle{plain}
\pagenumbering{arabic}

\section{Introduction} 

Strongly coupled quantum field theories play a central role
in the description of nature at the most fundamental level. Quantum
chromodynamics describes the observed strong interactions, and
other strongly interacting theories are part of many
efforts to extend the standard model.  But strongly coupled
field theories are notoriously difficult to analyze directly.
Because of this, general constraints on their behavior can be very
useful.

In a recent paper \cite{acs}, a new general constraint on strongly
coupled field theories was proposed. It takes the form of an
inequality limiting the number of massless degrees of freedom in
the infrared description of a field theory to be no larger than the
number of ultraviolet degrees of freedom. This inequality was 
conjectured to apply to all asymptotically
free theories (those governed by a free ultraviolet fixed point),
whose infrared behavior is also governed by a fixed point, not
necessarily free. It was noted that the inequality can also apply
to certain theories with interacting ultraviolet fixed points.

 The inequality is formulated using finite temperature as a device
to probe all energy scales, with the degree-of-freedom count
defined using the free energy of the field theory. The
zero-temperature theory of interest is characterized using the
quantity $f_{IR}$, related to the free energy by
\begin{equation}
\label{eq:firdef}
  f_{IR} \equiv -\frac{90}{\pi^2} \lim_{T\to 0} \frac{\cal F}{T^4} \ , 
\end{equation}
where $T$ is the temperature and $\cal F$ is the conventionally
defined free energy per unit volume (which is equal to minus the
pressure). This limit will be well defined if the theory has an
infrared fixed point. For the special case of an infrared-free
theory, $f_{IR}$ is simply the number of massless bosonic degrees
of freedom plus $7/8$ times the number of massless fermionic
degrees of freedom. The corresponding expression in the large $T$
limit is
\beq
  \label{eq:fuvdef}
  f_{UV} \equiv -  \frac{90}{\pi^2}\lim_{T\to \infty} \frac{\cal F}{T^4} \ .
\eeq
Just as in the infrared, this limit will be well defined if the
theory has an ultraviolet fixed point. For an asymptotically free
theory, $f_{UV}$ counts the underlying, ultraviolet degrees of
freedom in a similar way.

In terms of these quantities, the conjectured inequality for
asymptotically-free theories is
\beq
  \label{eq:ineq}
  f_{IR} \le f_{UV} \ .
\eeq
In Ref. \cite{acs}, the inequality was compared to known results
and then used to derive new results for several strongly coupled,
vector-like gauge theories. \footnote{\footnotesize{While the fixed
points are necessary to ensure the existence of the limits Eqs.
(\ref{eq:firdef},\ref{eq:fuvdef}), one can imagine applying the
inequality when the ultraviolet fixed point is only approximate,
that is when it is a feature of an effective low energy theory.
$f_{UV}$ would then count degrees of freedom for energy scales much
larger than masses and confinement or symmetry breaking scales in
the effective theory, but much smaller than the scale of new
physics.}}

 In the present paper, we extend this study to chiral gauge theories. These
theories, in which the left- and right- handed fermions couple differently to
the gauge fields, are potentially important examples of strongly coupled gauge
field theories.  The motivation for the effort to understand better the
nonperturbative behavior of chiral gauge theories stems not just from their
field-theoretic interest, but also from their possible application to
physics beyond the standard model, including (i) models of particle
substructure that can produce massless composite fermions and hence account for
the fact that the observed fermions have masses much smaller than the lower
bound on a hypothetical compositeness scale, and (ii) dynamical symmetry
breaking of electroweak and higher symmetries.  A key feature is that the
fermion content is subject to an additional constraint not present in vectorial
gauge theories, namely the absence of gauge and global $\pi_4$ (Witten)
anomalies (and the absence of mixed gauge-gravitational anomalies, if one
includes gravity). This greatly reduces the number of models that one can
consider. On the other hand, a number of powerful techniques that one can use
for vectorial gauge theories are absent for chiral gauge theories; these
include correlation function inequalities (since the fermion measure is not
positive).  There are also complications in trying to formulate chiral gauge
theories on the lattice because of fermion doubling.  Nevertheless, one can
still use the 't Hooft global anomaly matching conditions \cite{thooft} as well
as large-$N$ methods.

 For asymptotically free chiral theories, a variety of infrared phenomena can
be consistent with global anomaly matching.  One possibility is that as in QCD
the gauge symmetry remains intact but the theory confines and the global
flavor symmetries break spontaneously. Another possibility is that confinement
sets in but that the global symmetries are unbroken.  This is realized by the
formation of gauge singlet, massless composite fermions, along with other
possible degrees of freedom. It is also possible that the theory does not
confine but is governed in the infrared by an interacting fixed point, possibly
weak. The symmetries will again remain unbroken.  Yet another well-known
possibility is that these theories dynamically break their own gauge
symmetries, e.g., by the formation of fermion condensates.  Each of these
possibilities except the last will play a role here.

We examine several anomaly-free chiral gauge theories.  Some of
these are automatically asymptotically free. For the others, we
restrict the number of fermion representations so that they are
asymptotically free. The quantity $f_{UV}$ may then be computed
perturbatively. We consider the possible infrared realizations of
the theories as allowed by global anomaly matching and large-$N$
methods, and compute $f_{IR}$. For some models, the inequality is
automatically satisfied, while for others the assumption of the validity of the
inequality provides a strong restriction on the infrared realization.

\section{SU($N$) Models}

We begin with models based on the gauge group SU($N$) \cite{drs,eppz}.
For all the models considered in this paper, the beta function is
generically written as
\beq
\beta = \mu \frac{d \alpha}{d\mu} = -\beta_1 \Bigl ( \frac{\alpha^2}{2\pi}
\Bigr ) - \beta_2 \Bigl ( \frac{\alpha^3}{4\pi^2} \Bigr ) + O(\alpha^4) \ , 
\label{betafun}
\eeq
where the terms of order $\alpha^4$ and higher are scheme-dependent.

The first model contains massless fermions transforming according to (i) a
symmetric tensor representation of SU($N$), $S = \psi^{(ab)}_L$, and (ii) $N+4$
conjugate fundamental representations $F^c_{iL} =\psi^a_{iL}$, where $a,b$ are
SU($N$) group indices, $i=1,..,N+4$ is a flavor index, and all fermion fields
are written as left-handed Weyl fields.  The $\beta$-function
coefficients are $\beta_1 = 3N-2$, so that the theory is asymptotically free,
and $\beta_2 = (1/4)(13N^2 - 30N + 1 + 12N^{-1})$. Since the theory is
asymptotically free, $f_{UV}$ is given by the free-field count of the
thermodynamic degrees of freedom:
\beq
f_{UV} = 2(N^2-1) + (7/4) [ (1/2)N(N+1) + (N+4)N ] \ . 
\label{finfsym}
\eeq

To determine $f_{IR}$, we assume that the theory confines and apply
the 't Hooft anomaly matching conditions.  The global flavor
symmetry group is $G_f =$ SU($N+4$) $\times$ U(1), where the
SU($N+4$) mixes the $(N+4)$ $F^c_i$ fields and the U(1) is the
linear combination of the original U(1)'s generated by $S \to
e^{i\theta_S}S$ and $F^c \to e^{i\theta_{F^c}}F^c$ that is left
invariant by instantons.

The anomaly matching conditions are
consistent with the hypothesis that the global flavor symmetry
 group $G_f$ is unbroken and the massless spectrum
 is comprised of gauge-singlet composite fermions transforming
  according to the antisymmetric second-rank tensor representation of $G_f$.
In the large-$N$ limit, it has in fact been argued \cite{eppz} that
this {\it is} the infrared spectrum of the theory.  From this
spectrum one can determine that
\beq
f_{IR} = \frac{7}{4} \frac{(N+4)(N+3)}{2} \ .
\label{fzerosym}
\eeq
whence
\beq
\Delta f \equiv f_{UV} - f_{IR} = (1/4)[ 15N^2 + 7N - 50 ] \ ,
\label{deltafsym}
\eeq
which is positive for all $N \ge 2$. The inequality is
automatically satisfied.

The second model contains massless fermions in (i) the antisymmetric tensor
representation of SU($N$), $A = \psi^{[ab]}_L$ and, if $N \ge 5$, also
(ii) $N-4$ conjugate fundamental representations
$F^c_{iL} =\psi^a_{iL}$, where $i=1,...,N-4$.
For the beta function, we have $\beta_1=3N+2 > 0$, and
$\beta_2 = (1/4)(13N^2 + 30N + 1 - 12N^{-1})$.  Since the theory is
asymptotically free, we find
\beq
f_{UV} = 2(N^2-1) + \frac{7}{4}\biggl [
\frac{N(N-1)}{2} + (N-4)N \biggr ] \ . 
\label{finfantisym}
\eeq
For $N \ge 6$, the global symmetry group is $G_f =$ SU($N-4$) $\times$
U(1), where the SU($N-4$) mixes the $N-4$ fields $F^c_i$ and the U(1) is the
linear combination of the original U(1)'s generated by $A \to e^{i\theta_A}A$
and $F^c \to e^{i\theta_F^c}F^c$ that is left invariant by instantons.  The
anomaly matching conditions are consistent with the conclusion that the
massless spectrum consists of gauge-singlet composite fermions transforming
according to the symmetric second-rank tensor representation of $G_f$. (In the
degenerate case, $N=4$, there are no massless fermions.)  From this spectrum,
we deduce that
\beq
f_{IR} = \frac{7}{4} \frac{(N-4)(N-3)}{2} \ .
\label{fzeroantisym}
\eeq
Whence
\beq
\Delta f = (1/4)[ 15N^2 - 7N - 50 ] \ ,
\label{deltafantisym}
\eeq
which is positive for the relevant range $N \ge 4$. The inequality
is again satisfied.

The third model is an extension of model 1 with the same gauge group
and fermions transforming as (i) a symmetric tensor representation
$S = \psi^{(ab)}_L$; (ii) $N+4$ conjugate fundamental
representations: $F^c_i =
\psi^c_{iL}$, where $i=1,...,N+4$; and (iii) $p$ pairs of fundamental and
conjugate fundamental representations $F_{iL}, \ F^c_{iL}, i=1,...,p$.

We have $\beta_1=3N-2-(2/3)p$ and $\beta_2=(1/4)\{13N^2-30N+1+12/N
-2p((13/3)N-1/N)\}$.  Hence, the theory is asymptotically free if
\beq
p < (9/2)N - 3 \ . 
\label{afby}
\eeq
We shall restrict $p$ so that this condition is satisfied. We then find
\beq
f_{UV} = 2(N^2-1) + \frac{7}{4}\biggl [ \frac{N(N+1)}{2}
+ (N+4)N + 2pN \biggr ] \ . 
\label{finfby}
\eeq

The global symmetry group is
\beq
G_f = SU(r) \times SU(p) \times U(1) \times U(1)' 
\label{gglobal3}
\eeq
where
\beq
r = N+4+p \ . 
\label{rdef}
\eeq
The first U(1) in (\ref{gglobal3}) is generated by $F_{iL} \to
e^{i\omega}F_{iL}$ and $F^c_{iL} \to e^{-i\omega}F^c_{iL}$, which
is a vectorial symmetry and hence is not affected by instantons; the
U(1)$^{\prime}$ is the one left invariant by instantons.  The anomaly
matching conditions are consistent with the suggestion \cite{by}
that the spectrum of the model consists of gauge-singlet massless
composite fermions transforming according to the representations
\beq
(\ \asym \ , \ 1) + (\ \overline{\fund} \ , \ \fund \ ) + (1, \
\overline{\sym} \ )
\label{model3comp}
\eeq
of $SU(r) \times SU(p)$ in $G_f$.

For simplicity, we will restrict our discussion to the large-$N$
limit. It was noted in Ref. \cite{eppz} that the above
composite-fermion realization of $G_f$ is not possible if this
limit is taken with fixed $p$. Assuming confinement, $G_f$ would
have to be broken to a smaller group. If, on the other hand, we
take the limit
\beq
N \rightarrow \infty, \quad p \rightarrow \infty, \quad \frac{p}{N} \sim O(1)
\label{pnlimit}
\eeq
(so that loops of the $p$ fermions are not suppressed), the above
realization with unbroken $G_f$ is possible.

  We therefore assume the limits in Eq. (\ref{pnlimit}), with
\beq
\frac{p}{N} \equiv \lambda < \frac{9}{2}
\label{pnaf}
\eeq
for asymptotic freedom, and calculate that
\beq
f_{IR} = \frac{7}{4}\biggl [ \frac{r(r-1)}{2} + rp + \frac{p(p+1)}{2}
\biggr ] \ , 
\label{fzeroby}
\eeq
where only the leading terms in the limit (\ref{pnlimit}) are
relevant. It follows that, keeping only these leading terms,
\beq
\Delta f = (15/4)N^2 - (7/2)p^2 \ . 
\label{deltafby}
\eeq
Hence, in the large-$N$ limit, $\Delta f > 0$ only if $\lambda$ (is finite and)
satisfies
\beq
\lambda \le \Bigl (\frac{15}{4} \Bigr )^{1/2} = 1.04 \ . 
\label{pnineq}
\eeq
which is a considerably stronger upper bound on $p$ than (\ref{pnaf}).
(For very small $\lambda \sim const./N$, we have already noted that
this realization is impossible and that $G_f$ must break.) Thus
there is a finite $\lambda$ range leading to composite fermions and
an unbroken $G_f$.

For larger $\lambda$ values, up to the asymptotic freedom limit
$\lambda = 9/2$, we expect the theory to be in the nonabelian
Coulomb phase, with an interacting infrared fixed point. Near the
upper end, the fixed point will be weakly interacting as determined
by the first two terms in the $\beta$ function. The fixed point is
then given by
 $\alpha_* = -2\pi \beta_1/\beta_2$, and since $\beta_{1} >0$, it exists
only if $\beta_{2} <0$. In
  the large-$N$ limit, one finds that the two-loop fixed point exists if
\beq
\frac{3}{2} < \lambda < \frac{9}{2} \ ,
\label{lamir}
\eeq
and its value is
\beq
\alpha_*N = \frac{8\pi(9-2\lambda)}{13(2\lambda-3)} \ .
\label{anir}
\eeq
Of course, the lower end of this range is not reliable since the
fixed-point coupling, as determined by the two-loop $\beta$ function,
approaches infinity. For $\lambda$ near 9/2, the infrared and
ultraviolet degrees of freedom are the same, and $\Delta f$ is
positive due to the (negative) perturbative correction to $f_{IR}$.
Whether the nonabelian Coulomb phase persists down to the value in
Eq. (\ref{pnineq}), $\lambda=1.04$, or even lower, is unknown. The
inequality says simply that the confined phase with massless
composite fermions and unbroken global symmetry group $G_f$ cannot
exist if $\lambda > 1.04$. The full exploration of the phases of
this model as a function of $\lambda$ is an interesting and
unsolved strong-coupling problem.

\section{Direct Product Gauge Groups}

We next examine a class of models first discussed by Georgi \cite{g} with
direct product gauge groups $G_k$ composed by alternating SU($N$) and SU($M$)
$k$ times, so that $G_2=$ SU($N) \times$ SU($M$), $G_3=$ SU($N) \times$ SU($M)
\times$ SU($N$), etc.   Letting $M(N,1)$ denote $M$ copies of the
representation $(N,1)$, we take the fermion content to be
\beq
M(N,1), \quad (\bar N, \bar M), \quad N(1,M) \quad {\rm for} \quad k=2 \ , 
\label{fermionsk2}
\eeq
\beq
M(N,1,1), \quad (\bar N, \bar M,1), \quad (1,M,N), \quad M(1,1,\bar N) \quad
{\rm for} \quad k=3 \ , 
\label{fermionsk3}
\eeq
\beq
M(N,1,1,1), \ (\bar N,\bar M,1,1), \ (1,M,N,1), \ (1,1,\bar N,\bar M), \
N(1,1,1,M) \ \ {\rm for} \ \ k=4 \ , 
\label{fermionsk4}
\eeq
etc.  For the $i$'th SU($N$),
$\beta(g_i) = -(g_i^3/(48 \pi^2))(11N-2M) + O(g_i^5)$ and for the
$j$'th SU($M$),
$\beta(g_j) = -(g_j^3/(48 \pi^2))(11M-2N) + O(g_j^5)$, so that the theory
is asymptotically free if
\beq
\frac{2}{11} < \frac{N}{M} < \frac{11}{2} \ , 
\label{nm411}
\eeq
which will be assumed henceforth.  Then
\beq
f_{UV} = 2[\ell_k(N^2-1)+\ell(M^2-1)]+(7/4)(k+1)MN \ , 
\label{finfg}
\eeq
where $\ell_k=\ell$ if $k=2\ell$ is even and $\ell+1$ if $k=2\ell+1$ is odd.

Again assuming confinement so that the massless, physical states
transform as singlets under $G_k$, one can apply the 't Hooft
anomaly matching conditions. Consider first even $k$; then the
global flavor symmetry group is $G_{f,k \ {\rm even}}$=SU($M)
\times$ SU($N) \times$ U(1), where the SU($M$) mixes the $M$ copies
of $(N,1,...,1)$, the SU($N$) mixes the $N$ copies of $(1,...,M)$,
and the U(1) is the one left invariant by instantons. The anomaly
matching conditions for the SU($N$)$^3$, SU($M$)$^3$,
SU($M$)$^2$U(1), and SU($N$)$^2$U(1) anomalies are consistent with
the conclusion, also supported by large $N,M$ arguments \cite{eppz}
(with $N/M$ fixed in the interval (\ref{nm411})) that the spectrum
consists of massless composite fermions that transform according to
the $(M,N)$ representation of $G_f$. Hence, $f_{IR, \ k \ {\rm
even}} = (7/4)MN$ and
\beq
\Delta f_{k \ {\rm even}} = 2\ell(N^2+M^2-2)+(7/4)kMN > 0 \ . 
\label{delta_f_gkeven}
\eeq

Next consider odd $k$; then the global flavor symmetry group is $G_{f,k \ {\rm
odd}}$=SU($M) \times$ SU($M) \times$ U(1), where the two SU($M$)'s mix the $M$
copies of $(N,1,...,1)$ among themselves and the $M$ copies of $(1,...,N)$
among themselves, respectively, and the U(1) is as before, with appropriate
charges. Anomaly matching conditions and large $M,N$
arguments imply that $G_f$ is spontaneously broken to $G_{diag.}=$ SU($M)_V
\times$ U(1).  Hence, the massless spectrum consists of $M^2-1$ Goldstone
bosons, so that $f_{IR, \ k \ {\rm odd}} = M^2-1$ and
\beq
\Delta f_{k =2\ell+1} = 2(\ell+1)(N^2-1) + (2\ell-1)(M^2-1)
+(7/2)(\ell+1)MN > 0 \ . 
\label{delta_f_gkodd}
\eeq
Thus, the global symmetry is believed to remain unbroken with massless
composite fermions for even $k$, and is believed to break, producing Goldstone
bosons, for odd $k$. In both cases, $\Delta f > 0$. The inequality provides no
new information about these models.

\section{Supersymmetric Models}

There are a number of chiral ${\cal N}=1$ supersymmetric gauge theories for
which dual descriptions are known which are weakly coupled in the IR. It is a
nontrivial test of the inequality to check that $f_{IR}$ as computed from
these duals is indeed less than $f_{UV}$ . We have performed this check for 
several chiral supersymmetric theories and found the predictions of the
inequality to be in agreement with the conjectured duals in all cases.  The
theories we studied include (i) all so-called s-confining theories as listed in
Ref. \cite{css}; (ii) an $SO(10)$ theory with chiral superfields transforming
according to a spinor and $N$ vector representations \cite{ps}, and
(iii) an $SU(N)$ theory with chiral superfields comprised of a symmetric 
tensor $S$ and $N+4$ antifundamental representations $\bar Q$ \cite{ps2}.

The theory (iii) is the ${\cal N}=1$ supersymmetric version of the model we
discussed in the beginning of Section II (with no tree-level superpotential). 
This theory is asymptotically free for all $N$, and we obtain 
\beq 
f_{UV}=\frac{15}4\left(N^2-1+\frac12 N(N+1)+N(N+4)\right) \ .
\label{fuvps}
\eeq
In \cite{ps} a dual ${\cal N}=1$ supersymmetric description of the IR dynamics
of this theory was proposed. The dual theory is non-chiral and has gauge group 
$SO(8)$, $N+4$ copies of the vector representation of $SO(8)$, a
spinor of $SO(8)$, and $(N+4)(N+5)/2 +1$ gauge singlets. 
For $N \ge 13$ the dual is IR free and we can calculate $f_{IR}$ by simply
counting dual fields
\beq
f_{IR}=\frac{15}4\left(28+8(N+4)+8+\frac12(N+4)(N+5)+1\right) \ .
\eeq
The constraint $f_{IR}\le f_{UV}$ then becomes $N \ge 9$; thus the inequality
is satisfied for all numbers of flavors for which we can reliably compute the
quantities $f_{UV}$ and $f_{IR}$.

Although we have concentrated on chiral gauge theories in this paper, we add
some remarks here on a particular vectorial gauge theory for which no dual
description is known. As we will see, applying the conjectured inequality 
yields a new constraint on the IR spectrum of this theory. The model
is an ${\cal N}=1$ supersymmetric $SU(N)$ gauge theory with the same matter
content as ${\cal N}=2$ supersymmetric QCD: $F$ chiral superfields $Q_i$ and
$\bar Q_i$ transforming in the fundamental and antifundamental representations
of $SU(N)$ and one chiral superfield $A$ in the adjoint representation. The
difference with the ${\cal N} =2$ theory is that we do not include a tree level
superpotential term $W=g Q A \bar Q$.  Setting the superpotential coupling to
zero breaks the extended supersymmetry to ${\cal N}=1$ and the resulting theory
is much less well understood. Our goal is to investigate how the inequality
constrains the infrared dynamics of this theory.

For $F< 2N$ the theory is free in the ultraviolet, and therefore
\beq
f_{UV}=\frac{15}4 \left(2(N^2-1)+2FN\right) \ .
\eeq
A full description of the IR theory is not known, but a few general
statements can be made. The classical scalar potential of this
theory has a large number of flat directions; the theory has a
moduli space of inequivalent vacua. It is well known that the
classical moduli space can be parametrized in terms of the
independent gauge invariant polynomials which in this case are
\beq
T_k={\rm Tr} A^k , \ k=2,...,N \quad {\rm and}\quad M_i= Q A^i \bar Q\
,\ i=0,...,N-1 \ ,
\eeq
in addition to several baryonic gauge invariants. Here the
$T_k$ are singlets, whereas the $M_i$ have $F^2$ components.

No infrared-free description to this theory has been found, and there
are arguments \cite{kss} that the theory at the origin of moduli
space remains interacting and is conformal in the infrared for any
value of $0<F<2N$. Still, the inequality can provide useful
information, as we will now discuss.

An interesting picture for the infrared dynamics has been put forward in
\cite{lst}, based on a study of a closely related $Sp(2N)$ theory.  If one
assumes that the infrared fixed point corresponds to an interacting
superconformal theory, then the superconformal algebra must contain an
anomaly-free $U(1)_R$ symmetry. It follows from the superconformal algebra that
the scaling dimensions of all chiral composites are equal to $3/2$ times their
charge under the $U(1)_R$. Unfortunately, the R-symmetry of our theory is not
unique; there exists a one-parameter family of R-symmetries of which one is the
$U(1)_R$ in the superconformal algebra. However one can still derive relations
between the scaling dimensions of the chiral composites $M_i$ and $T_k$. The
result of such an analysis is that for $F < N/2$, the scaling dimensions of
some composites, as determined from the superconformal R-charges, drop below
the unitarity bound.  This then implies that these composites decouple from the
interacting conformal theory.  Similar behavior is anticipated for $N/2 \le F <
2N$.  This picture for the dynamics is also supported by duality \cite{lst}.
In summary, we expect that in the deep infrared this theory splits into
disjoint sectors: one sector with an interacting conformal theory, and other
sectors consisting only of a set of $m$ free composite fields $M_i$.

We now apply the inequality to get a constraint on this scenario.
To calculate $f_{IR}$ we add the contributions from each of
the disjoint sectors. We cannot calculate the contribution
from the conformal sector, but we know that it is positive.\footnote{
\footnotesize{It follows from the definition in Eq. (\ref{eq:firdef})
together with ${\cal F}=-p$, that $f_{IR}$ has the same sign as the
pressure, which is always positive.}}
This allows us to compute a lower bound on
$f_{IR}$ by adding only the contributions from the $m$ free fields
$M_0,..., M_{m-1}$:
\beq
f_{IR}\ge \frac{15}4 m F^2 \ .
\eeq
Demanding that $f_{IR}\le f_{UV}$ then gives $m\le 2(\frac{N^2}{F^2}+
\frac{N}F-\frac1{F^2})$. Since only the first $N$ of the $M_i$ are
independent gauge invariants, we already have that $m\le N$, so that
the inequality provides new information only for $F>2$. The
constraint then simplifies to
\beq
m < 2\frac{N}{F}(\frac{N}{F}+1) \ .
\eeq
We see that for larger $F$ this constraint can be very significant.
For example, for $F$ near the asymptotic freedom bound at $F=2N$ we
find $m<3/2$, implying that only the first in the series of mesons, $M_0$,
can decouple from the conformal theory and become a free field.

The case of $SU(2)$ is somewhat special and leads to a tight
constraint. For $SU(2)$ both $Q$ and $\bar Q$ transform in the fundamental
doublet representation of $SU(2)$, and the global flavor symmetry is enhanced
to $SU(2F)$. The mesons $M_{0,1}$ transform as tensors under this enlarged
flavor symmetry. $M_0$ is an antisymmetric tensor, whereas $M_1$ is
a symmetric tensor. The constraint then allows the following scenario:
for $F\ge4$ the theory is not asymptotically free and we do not get
a constraint; for $F=2,3$ only the meson $M_0$ can decouple and become
free; and for $F=1$ both $M_0$ and $M_1$ may be free.

\section{Summary}

In summary, we have performed a quantitative comparison of the
ultraviolet and infrared degrees of freedom, as measured by $\Delta
f=f_{UV}-f_{IR}$, for several chiral gauge theories and for one
vector-like, supersymmetric gauge theory. We have shown that in
some models $\Delta f$ is automatically positive, whereas in
others, such as the third model of Section II and the vector-like,
supersmmetric model ( Section IV), the conjectured inequality
$\Delta f \ge 0$ places interesting restrictions on the low energy
structure of the theory.

This research was partially supported by the DOE and NSF grants
DE-FG02-92ER-4074 (TA), DE-FG02-92ER-40704 (AC), DE-AC03-76SF00515 (MS), and
NSF-PHY-97-22101 (RS).  T. A. acknowledges the hospitality of the ITP, Stony
Brook during a visit in December, 1998, where this work was initiated.

\vfill
\eject
\end{document}